\documentstyle[12pt]{article}

\newlength{\dinwidth}
\newlength{\dinmargin}
\newcommand{\ba}{\begin{array}}
\newcommand{\ea}{\end{array}}
\newcommand{\be}{\begin{equation}}
\newcommand{\ee}{\end{equation}}
\newcommand{\bee}{\begin{eqnarray}}
\newcommand{\eee}{\end{eqnarray}}

\font\cmss = cmss12

\def\identity{{\rlap{1} \hskip 1.6pt \hbox{1}}}
\def\integer{{\rlap{\cmss Z} \hskip 1.8pt \hbox{\cmss Z}}}
\def\laplace{{\kern1pt\vbox{\hrule height 1.2pt\hbox{\vrule width 1.2pt\hskip
  3pt\vbox{\vskip 6pt}\hskip 3pt\vrule width 0.6pt}\hrule height 0.6pt}
  \kern1pt}}
\def\scriptlap{{\kern1pt\vbox{\hrule height 0.8pt\hbox{\vrule width 0.8pt
  \hskip2pt\vbox{\vskip 4pt}\hskip 2pt\vrule width 0.4pt}\hrule height 0.4pt}
  \kern1pt}}

\def\roughly#1{\raise.3ex\hbox{$#1$\kern-.75em\lower1ex\hbox{$\sim$}}}

\def\gym{g^2_{\scriptscriptstyle YM}}
\def\gymb{\bar{g}^2_{\scriptscriptstyle YM}}

\begin{document}

\thispagestyle{empty} \addtocounter{page}{-1} 
\begin{flushright}
SNUTP 97-140\\
{\tt hep-th/9710192}\\
\end{flushright}
\vspace*{1.3cm} \centerline{\Large {\bf Non-Orientable M(atrix) Theory
\footnote{{ 
{Work supported in part by the NSF-KOSEF Bilateral Grant, 
KOSEF SRC-Program, Ministry of Education Grant BSRI 97-2410,
and The Korea Foundation for Advanced Studies. }}}}} 
\vspace*{1.7cm} 
\centerline{\large {\bf Nakwoo Kim and Soo-Jong Rey}} 
\vspace*{1.5cm}
\centerline{\large {\it Physics Department \& Center for Theoretical Physics}}
\vskip0.2cm 
\centerline{\large {\it Seoul National University, Seoul 151-742 KOREA}} 
\vspace*{0.6cm}
\centerline{\large\tt nakwoo@fire.snu.ac.kr, sjrey@gravity.snu.ac.kr}
\vspace*{1.5cm} 
\centerline{\Large\bf abstract} \vskip0.6cm
M(atrix) theory description is investigated for M-theory compactified on
non-orientable manifolds. Relevant M(atrix) theory is obtained
by Fourier transformation in a way consistent with T-duality.
For nine-dimensional compactification on Klein bottle and M\"obius strip,
we show that M(atrix) theory is (2+1)-dimensional ${\cal N} = 8$ 
supersymmetric U(N) gauge theory defined on {\sl dual Klein bottle}
and {\sl dual M\"obius strip} parameter space respectively.
The latter requires a twisted sector consisting of sixteen chiral fermions 
localized parallel to the boundary of dual M\"obius strip and  
defines Narain moduli space of Chaudhuri-Hockney-Lykken (CHL) heterotic string. 
For six-dimensional CHL compactification $\left(({\bf S}_1/\integer_2) \otimes 
{\bf T}^4 \right)/ \Gamma_{\rm CHL}$ 
we show that low-energy dynamics of M(atrix) theory is described by 
(5+1)-dimensional ${\cal N} = 8$ supersymmetric
U(N)$\times$U(N) gauge theory defined on {\sl dual orbifold} parameter space 
of $(\widetilde {\bf S}_1 \otimes \widetilde{K3} )/\integer_2$. 
Spacetime spectrum is deduced from BPS gauge field configurations consistent 
with respective involutions and is shown to agree with results from M-theory 
analysis. 
\vspace*{1.1cm}

\setlength{\baselineskip}{18pt}
\setlength{\parskip}{12pt}

\newpage


\section{Introduction}

\setcounter{equation}{0} 
At present, M(atrix) theory~\cite{bfss} 
is the only available non-perturbative description of M-theory~\cite{witten}, the theory which unifies all known perturbative string theories. 
Based on light-front~\footnote{or light-cone~\cite{susskind, seibergnew}} 
Hamiltonian formalism and 
holographic principle, it has been identified that the fundamental degrees of 
freedom of M-theory are Dirichlet zero-branes (D0-branes) of Type IIA string. 
Dynamics of N D0-partons is described by ${\cal N}$= 16 supersymmetric U(N) 
Yang-Mills quantum mechanics and it is the latter that defines the M(atrix) 
theory.

One of the most pressing issues in M(atrix) theory is to identify proper
description of M-theory compactification. In the limit $d$-dimensional
compactified space shrinks to a vanishing size, it has been found that
Fourier transformation and T-duality yields the corresponding M(atrix)
theory is $(d+1)$-dimensional quantum field theory which reduces to 
a supersymmetric Yang-Mills theory at low energy.
Consider a compactified space ${\cal X}_d = {\bf R}_d / \Gamma$, a quotient
of d-dimensional flat space by a symmetry group $\Gamma$. M(atrix) theory
on ${\cal X}_d$ is obtained by taking into account of all image D0-partons
located at orbits of $\Gamma$. 
Dynamical fields of the M(atrix) theory are gauge potential, 
transverse coordinates $X^\mu \equiv (A_0, X^i), \quad (i=1, \cdots, 9)$,
and super-partner spinor ${\bf \Theta}^\alpha, \quad (\alpha = 1, \cdots,
16)$.
As the volume of ${\cal X}_d$ is shrunken to zero, the configurations are
most conveniently described in terms of T-dual picture. Under T-duality
winding excitations are mapped into momentum excitations. As such, sum
over winding string configurations is equivalent to Fourier transformation
in dual space $\widetilde{\cal M}$. For the simplest ${\bf
S}_1$ compactification, the transformation has been performed explicitly and
the result turns out to be large-N limit of (1+1)-dimensional ${\cal N} = 16$
supersymmetric U(N) gauge theory with one adjoint matter multiplet.

Also studied are M(atrix) theory description of M-theory compactified on
orbifolds. Among the more interesting ones are ${\bf S}_1/\integer_2$~\cite{kimrey1, kabatrey}, ${\bf T}^5/\integer_2$~\cite{kimrey2} and 
${\bf T}^9/\integer_2$~\cite{kimrey3} orbifolds. 
Interesting feature common to all of them 
is that corresponding M(atrix) gauge theories
are {\sl chiral}, hence, entails nontrivial dynamics not encountered in
maximally supersymmetric
toroidal compactifications. Charge conservation, supersymmetry and gauge
anomaly cancellations then introduce {\sl twisted sector} uniquely to the
M(atrix) theory. The twisted sectors all arise from fixed points of the
orbifold, which can be reduced or eliminated completely if one take further 
quotient by freely acting involutions. The simplest quotient
manifolds obtained by freely acting involution arise in two-dimensions:
Klein bottle and M\"obius strip. M-theory compactified on these manifolds
have been discussed briefly~\cite{chaupol, dabpark}. It has been identified
that M-theory compactified on Klein bottle is a projection of Type IA
string, while the one on M\"obius strip is Chaudhuri-Hockney-Lykken (CHL) 
string~\cite{chl} with $E_8$ or $SO(16)$
gauge groups.

In this paper, we construct M(atrix) theories for compactifications on
non-orientable manifolds via straightforward application of Fourier 
transformation and T-duality. As will be shown in sections 3 and 4, 
corresponding
M(atrix) theories are (2+1)-dimensional ${\cal N} = 8$ supersymmetric gauge
theories living on {\sl dual} Klein bottle or M\"obius strip parameter
spaces. These so-called non-orientable
M(atrix) theories are quotients of the M(atrix) theory compactified on 
${\bf T}_2$ by appropriate freely acting involutions, and are obtained 
straightforwardly from Fourier and T-duality tranformations.   
For both Klein bottle and M\"obius strip, the M(atrix) gauge
group is U(N)$\times$U(N) except that, at the boundary of M\"obius strip, 
the gauge group is promoted to SO(2N). 
To ensure local charge conservation and cancel supersymmetry and
gauge anomalies, a twisted sector consisting of sixteen (1+1)-dimensional
chiral fermions needs to be introduced. Deformation of chiral fermions away
from the orbifold boundary corresponds to turning on Wilson line in the CHL
heterotic
string. We also study BPS configurations of M(atrix) gauge theory
and deduce spacetime spectrum from them. We find a complete agreement with 
the spectrum deduced earlier from string theory consideration. 
Non-orientabl M(atrix) theories can
be straightforwardly generalized to compactification on other
higher-dimensional quotient space by freely acting involutions. In 
section 5, we illustrate this for toroidally compactified six-dimensional 
CHL string. We find that, at low-energy, the corresponding M(atrix) theory 
is supersymmetric gauge theory whose parameter space 
is an orbifold limit of ${\bf S}_1 \times K3$ modded out by $\integer_2$
automorphism of K3.
While this work is finished and is being written, we have received a 
work~\cite{zwart} that overlaps with parts of sections 3 and 4. We disagree, however, with part of the derivation and conclusion thereof.


\section{M(atrix) theory on Torus} 

We begin with recapitulating M(atrix) theory description for M-theory
compactified on a torus ${\bf T}_2$. It is intended to fix notations and 
essential aspects that will become relevant for later discussions. 
Consider M(atrix) theory for compactification on a rectangular torus
${\bf T}_2$ of size $(2 \pi R_1) \times (2 \pi R_2)$. The theory is defined 
in terms of $N$ D0-partons living on ${\bf T}_2 \times {\cal M}^\perp_8$. 
Large-$N$ limit $N \rightarrow \infty$ will be implicit throughout. 
Dynamics among D0-partons is described completely once all allowed 
configurations of open fundamental string connecting the partons are 
specified. They are encoded into $\left( N \times N \right)$ 
M(atrix) fields $X^\mu \equiv (A_0, X^i)$, where $A_0(t)$ denotes gauge 
potential and $X^i(t) \quad (i=1, \cdots, 9)$ transverse position matrices. 
To account for the parton dynamics correctly, it is necessary to 
take into consideration of string configurations that wind around one-cycles 
of ${\bf T}_2$.  This is achieved most conveniently by arraying image 
D0-partons at orbits of the symmetry group 
$\Gamma = \integer \otimes \integer$ on 
the covering space ${\bf R}_2$, viz. ${\bf T}_2 = {\bf R}_2 / 
(\integer \otimes \integer)$. 
Located on each fundamental cell, which is labelled by an indicial vector 
${\bf k} \equiv (k_1,k_2), \quad k_1, k_2 = \cdots, -2, -1, 0, 1, 2, \cdots $, 
are exactly N D0-partons. Introduce covering space M(atrix) fields 
$X^\mu_{{\bf k}, {\bf l}}$ denoting interactions along $\mu$-direction 
between N D0-partons in 
${\bf k} = (k_1,k_2)$-th and N D0-partons in ${\bf l} = (l_1, l_2)$-th cells. 
The cell indices ${\bf k, l}$ are treated as generalized row and column 
indices.  
Thus $X^\mu_{\bf k, l}$ for each $\bf k, l$ are $(N \times N)$ sub-matrices of 
infinite-dimensional matrices.
Let also ${\bf e}_1 \equiv (1,0), \quad {\bf e}_2 \equiv (0, 1)$ denote 
action of $\Gamma$ on the covering space along each direction and
$G_{ij} = {\rm diag.} (R_1, R_2)$ metric of ${\bf T}_2$.
Modding out by the symmetry group $\Gamma = \integer \times \integer$ 
imposes the following periodicity conditions:
\bee
{X^1}_{{\bf k} + {\bf e}_1, \,\, {\bf k} + {\bf e}_1} &=& 
{X^1}_{\bf k, \, k} 
+ {\bf n} \cdot ( 2 \pi G ) \cdot {\bf e}_1 \,\, {\bf I}_{N \times N}
\nonumber \\
{X^2}_{{\bf k} + {\bf e}_2, \,\, {\bf k} + {\bf e}_2 } &=& 
{X^2}_{\bf k, \, k} + {\bf n} \cdot (2 \pi G ) \cdot {\bf e}_2 \, \, 
{\bf I}_{N \times N}
\nonumber \\
{X^\mu}_{{\bf k} + {\bf e}_1, \,\, {\bf l}+{\bf e}_1} &=&   
{X^\mu}_{{\bf k} + {\bf e}_2, \,\, {\bf l} + {\bf e}_2 } = 
{X^\mu}_{\bf k, \, l} \quad, \hskip0.5cm ({\bf k} \ne {\bf l})
\label{toruscp}
\eee
where ${\bf n} = (n_1, n_2)$ are integer-valued lattice vector.
By successive operation of lattice shift, it is possible to bring the
configuration at $({\bf k, l})$ to the one at $({\bf k - l, 0})$. Thus, 
we define $X^\mu({\bf k - l}) \equiv X^\mu_{{\bf k-l}, \, {\bf 0}}$
and express the periodicity condition Eq.(~\ref{toruscp}) compactly in 
terms of Fourier transformed fields defined on two-dimensional parameter 
space ${\bf y} \equiv (y_1, y_2)$:
\be
X^\mu({\bf y}, t) = \sum_{{\bf k}}
{X^\mu} ({\bf k}, t) \, \exp \Big[i {\bf k} \cdot {\widetilde G} 
\cdot {\bf y} \Big], \hskip1cm (i=1, \cdots, 9)
\label{fouriertransf}
\ee
where
\bee
\widetilde{ G}_{ij} &=& {\rm diag.} \left(
\widetilde{ R}_1, \widetilde{ R}_2 \right),
\nonumber \\
&& \widetilde{ R}_{1,2} = {\ell_{11}^3  / R_{11} \, R_{1,2} }.
\eee
To emphasize distinctive role of compactified and non-compact spacetime 
coordinates, we introduce (2+1)-dimensional fields
$A_\alpha = (A_0, A_1 \equiv X^1, A_2 \equiv X^2)$ and
$Y^i = X^i \, (i = 3, \cdots, 9)$.
Then, the Chan-Paton condition Eq.(\ref{toruscp}) turns into boundary
conditon of the (2+1)-dimensional fields: 
\bee
A_\alpha ({\bf y}, t) &=&
A_\alpha ({\bf y}  + 2 \pi {\widetilde G} \cdot {\bf n}, \,\, t)
\nonumber \\
Y^i
({\bf y}, t) &=& 
Y^i \, (
{\bf y} + 2 \pi {\widetilde G} \cdot {\bf n}, \,\, t),
\hskip1cm {\bf n} = (n_1, n_2) \in \integer \times \integer.
\eee
After Fourier transformation is done explicitly, the M(atrix) quantum
mechanics turns into (2+1)-dimensional ${\cal N} = 16$ supersymmetric $U(N)$ 
gauge theory with one adjoint matter multiplet, which is equivalent to 
dimensional reduction of (9+1)-dimensional supersymmetric U(N) Yang-Mills
theory to (2+1)-dimensions. The transverse rotational invariance is realized
as $Spin_R(7)$ R-symmetry. Lagrangian of the gauge theory is given by
\begin{eqnarray}
L & = & - {\frac{1 }{2 g^2_{\scriptscriptstyle YM}}} \int_{{\widetilde T}_2}
d^2 {\bf y} \, 
{\rm Tr} \Big\lbrace F_{\alpha \beta} F^{\alpha \beta} + 2 D_\alpha
Y_i D^\alpha Y^i - [Y_i, Y_j][Y^i, Y^j]  \nonumber \\
& & \qquad\qquad\qquad\qquad - 2 i {\overline \psi}_A \gamma^\alpha D_\alpha
\psi_{A} + 2 i {\overline \psi}_A \gamma^i_{AB} [Y_i, \psi_B] \,\, %
\Big\rbrace  \label{untwisted}
\end{eqnarray}
Denoting the M-direction radius $R_{11}$, the gauge coupling is related
to the parameters of the compactification torus ${\bf T}_2$ as  
\be
\gym = {2 R_{11} \over R_1 R_2}.
\ee
Our notations are as follows. The SO(9,1) gamma matrices in Majorana-Weyl 
representation are decomposed into tensor product of SO(2,1)$\times$SO(7) 
gamma matrices:
\begin{equation}
\Gamma^\alpha = \left[
\begin{array}{cc}
0 & \gamma^\alpha \otimes {\rlap{1} \hskip 1.6pt \hbox{1}}_{8 \times 8} \\
\gamma^\alpha \otimes {\rlap{1} \hskip 1.6pt \hbox{1}}_{8 \times 8} & 0
\end{array}
\right] \hskip1cm \Gamma^i = \left[
\begin{array}{cc}
0 & -i {\rlap{1} \hskip 1.6pt \hbox{1}}_{2 \times 2} \otimes \gamma^i \\
+ i {\rlap{1} \hskip 1.6pt \hbox{1}}_{2 \times 2} \otimes \gamma^i & 0
\end{array}
\right]
\end{equation}
such that 
\begin{equation}
\Gamma^{(11)} \equiv \Gamma^0 \cdots \Gamma^9 = \left[
\begin{array}{cc}
+ {\rlap{1} \hskip 1.6pt \hbox{1}} & 0 \\
0 & - {\rlap{1} \hskip 1.6pt \hbox{1}}
\end{array}
\right].
\end{equation}
All the spinors $\psi_A$ satisfy $\Gamma^{(11)} \psi_A = + \psi_A$.
The (2+1)-dimensional gamma matrices are denoted by  $\gamma^\alpha$:
\begin{equation}
\gamma^0 = \pmatrix{0 & -i \cr +i & 0} \;\;\; \gamma^1 =
\pmatrix{0 & +i \cr +i & 0} \;\;\; \gamma^2 = %
\pmatrix{+i & 0 \cr 0 & -i}
\end{equation}

Our notations are as follows. The thirty-two supersymmetries of M-theory
can be decomposed as ${\bf 16}_+ \oplus {\bf 16}_-$ of SO(9,1).
Once the infinite momentum frame is chosen, the supersymmetries in the
${\bf 16}_-$ are broken and become non-linearly realized kinematical
supersymmetries of M(atrix) theory. The other half, ${\bf 16}_+$
supersymmetries are unbroken and defines the dynamical supersymmetries
of M(atrix) theory.
Then, the ${\bf 16}_\pm$ spinors associated with dynamical and kinematical
supersymmetries are reduced to (2+1)-dimensional spinors $\epsilon_A$ and 
$\eta_A$ that are  
taken to be in inequivalent, opposite-sign representations of the 
(2+1)-dimensional Clifford algebra. 

The M(atrix) gauge theory possesses thirty-two supercharges, among which half 
of them are dynamical supersymmetries in the light-front kinematics: 
\begin{eqnarray}
\delta_\epsilon A_\alpha &=& + {\frac{i }{2}} \, {\overline \epsilon}_A
\gamma_\alpha \psi_A  \nonumber \\
\delta_\epsilon X^i &=& -{\frac{1 }{2}} \, {\overline \epsilon}_A
\gamma^i_{AB} \psi_B  \nonumber \\
\delta_\epsilon \psi_A &=& - {\frac{1 }{4}} F_{\alpha \beta} \gamma^{\alpha
\beta} \epsilon_A - {\frac{i }{2}} D_\alpha X_i \gamma^\alpha \gamma^i_{AB}
\epsilon_B - {\frac{i }{4}} [X_i, X_j] \gamma^{ij}_{AB} \epsilon_B \,.
\label{dynamicsusy}
\end{eqnarray}
The other half are  kinematical supersymmetries 
\begin{eqnarray}
\delta_\eta A_\alpha &=& \delta_\eta X_i = 0  \nonumber \\
\delta_\eta \psi_A &=& \eta_A {\rlap{1} \hskip 1.6pt \hbox{1}}
\label{kinematicsusy}
\end{eqnarray}
and acts only on the center-of-mass U(1)$\subset$U(N).
It is evident that the M(atrix) theory has $Spin(7)_R$ R-symmetry, which 
in fact encodes the rotational invariance on the transverse noncompact space.

\section{M(atrix) Theory on Klein bottle}
Consider nine-dimensional M-theory compactified on a Klein bottle
${\bf K}_2$. Klein bottle ${\bf K}_2$ of area
$(\pi R_1) \times (2 \pi R_2)$ is obtained as a quotient of torus ${\bf T}_2$
of area $(2 \pi R_1) \times (2 \pi R_2)$ by symmetry group
$\Gamma_{\bf K}$:
\be
\Gamma_{\bf K} \quad : \quad {\bf x} \rightarrow \hat{\bf x} + \pi G \cdot
{\bf e} \quad, \quad \hat{\bf x} \equiv (- x^1, x^2)
\quad \quad
{\bf e} \equiv ({\bf e}^1, {\bf e}^2),
\ee
viz. parity along $1$-direction accompanied by half-period shifts in both 
directions.  To define the corresponding M(atrix) theory, we put N 
D0-partons on ${\bf K}_2$ and study their dynamics. From the
covering space ${\bf T}_2$ point of view, this is to place 2N D0-partons:
original N D0-partons and image N D0-partons. They form a single $\integer_2$
orbit of the symmetry group $\Gamma_{\bf K}$. In what follows D0-parton
dynamics on the ${\bf K}_2$ will be referred as \sl Klein bottle M(atrix)
theory. \rm

\subsection{Chan-Paton Condition}
In this section, we will find prescription of Klein bottle M(atrix) theory
in two steps. We first study 2N D0-parton dynamics on a single fundamental cell
of ${\bf T}_2$. Subsequently, utilizing the result in section 2, we obtaine 
the Klein bottle M(atrix) theory in terms of ${\bf R}_2$ covering space of 
${\bf T}_2$ with an appropriate projection, viz. (2+1)-dimensional gauge 
theory on a dual parameter space (which will be determined in due course of
foregoing construction).

On a single fundamental cell of ${\bf T}_2$, infinitesimally short open string 
configurations among 2N D0-partons define parton dynamics and are denoted
by $(2N \times 2N)$ matrix fields $X^\mu \equiv \left( A_0, \,\, X^i \right)
\quad \quad (i=1, \cdots, 9)$.
Compactification on Klein bottle ${\bf K}_2$ is then realized by imposing
$\Gamma_{\bf K}$ projection on the D0-parton configurations, hence, on these
matrix fields. In the previous work~\cite{kimrey1}, we have found that the
projection is given in terms of {\sl local} Chan-Paton conditions on each
fundamental cell of ${\bf T}_2$:
\begin{eqnarray}
A_0 &=& - M \cdot {A_0}^{\rm T} \cdot M^{-1} \nonumber \\
X^1 &=&-M \cdot X^{1{\rm T}} \cdot M^{-1}  - \left( \pi \cdot
G \right) \cdot {\bf e}_1 \, {\bf I}_{N \times N} \otimes \sigma^3 \nonumber \\
X^2 &=&+M \cdot X^{2{\rm T}} \cdot M^{-1}
- \left( \pi \cdot G \right) \cdot {\bf e}_2 \, {\bf I}_{N \times N} 
\otimes \sigma^3, \nonumber \\
X^I &=& + M \cdot X^{I {\rm T}} \cdot M^{-1} \quad,
\quad \quad \quad \quad \left( I = 3, \cdots, 9 \right) ,
\label{originalkbmatrix}
\end{eqnarray}
where $G_{ij} = {\rm diag.} \left( R_1, R_2 \right)$ denotes the metric of covering space ${\bf T}_2$ and
$M={\bf I} \otimes \sigma^1,\,\, {\bf I}\otimes \sigma^2$ as was determined
~\footnote{
Equivalently, the matrices 
along the Klein bottle can be parametrized as
\be
X^1 = \pmatrix{ X & U  \cr V & -X^{\rm T} } \quad , \quad
X^2 = \pmatrix{ Y + \pi R_2  {\bf I} & W \cr Z & +Y^{\rm T} }. 
\ee
} in the previous work~\cite{kimrey1}.
We have taken transpose action between the original and the image D0-partons 
since, in the Type IIA string limit, the Klein bottle corresponds to 
orientifold and gives rise to unoriented membrane. 
Note that Chan-Paton condition should include a condition on M(atrix) gauge 
potential $A_0$ as well in order to maintain the mapping as a symmetry of
M(atrix) quantum mechanics, by-now a well understood fact.

From M(atrix) gauge symmetry point of view, what does the inhomogenous term 
proportional to ${\bf I}_{N \times N} \otimes \sigma^3$ 
in Eq.(\ref{originalkbmatrix}) signify? To understand this note that, 
for the above choices of $M$, 
\begin{equation}
\pi R\pmatrix{ {\bf I} & 0 \cr 0 & -{\bf I}}=
\frac 12\pi R\pmatrix{{\bf I} & 0 \cr 0 & -{\bf I}}%
-\frac 12\pi RM \cdot \pmatrix{+{\bf I} & 0 \cr 0 & - {\bf I}} \cdot M^{-1}
\end{equation}
Hence, the Chan-Paton conditions for $X^{1,2}$ can be re-expressed as  
\bee
\left( X^1 + {1 \over 2} \pi R_1 {\bf I} \right) &=& - \, M \cdot 
\left( X^1 + {1 \over 2} \pi R_2 {\bf I} \right)^{\rm T} \cdot M^{-1} 
\nonumber \\
\left( X^2+\frac 12\pi R_2 {\bf I} \right) &=& + \, M \cdot \left( 
X^2+\frac 12\pi R_2 {\bf I} \right)^{\rm T} \cdot M^{-1}.
\eee
This shows that along $1,2$-directions, there are nontrivial Wilson line  
backgrounds but the D0-parton coordinates measured {\sl relative to these 
Wilson line} satisfy {\sl homogeneous} Chan-Paton condition. 

We now extend the Chan-Paton condition to other image cells in the covering
space.
Following exactly the same procedure as in Section 2 for ${\bf T}_2$, we
introduce cell-indexed coordinate matrices $X^\mu_{\bf k, l}$, 
which are (2N$\times$ 2N) matrices for each $\bf k,l$. 
The matrices in Eq.(\ref{originalkbmatrix}) are 
diagonal sub-blocks $({\bf k} = {\bf l})$ of these infinite-dimensional 
matrices. We now extend the Chan-Paton condition Eq.~(\ref{originalkbmatrix}) 
to all image cells in ${\bf R}_2$ covering space and rewrite them in terms 
of $X^\mu_{\bf k,l}$ matrices.  In particular, the diagonal, $2N$-dimensional 
sub-blocks of the extended condition should yield the same condition as 
Eq.~(\ref{originalkbmatrix}). With such proviso, the extended Chan-Paton
condition is given by  
\begin{eqnarray}
A_{0 \bf k, l} &=&- \, {\cal M}_{{\bf k}, {\bf k}'} \,\left( \, M
\cdot {{A_0}^{\rm T}}_{{\bf l}', {\bf k}'}
\cdot M^{-1} \, \right)\, {{\cal M}^{-1}}_{\bf l', l}
\nonumber \\
{X^1}_{\bf k, l} &=&- \, {\cal M}_{{\bf k}, {\bf k}' } \, \left( \, M
\cdot {X^{1 {\rm T}}}_{ {\bf l}', {\bf k}' }
\cdot M^{-1} \, \right) \, {{\cal M}^{-1}}_{ \bf l', l}
\nonumber \\
{X^2}_{\bf k, l} &=&+{\cal M}_{{\bf k}, {\bf k}' } \, \left( \, M
\cdot {X^{2 {\rm T}}}_{  {\bf l}', {\bf k}' } 
\cdot M^{-1} \, \right) \, { {\cal M}^{-1} }_{ \bf l',l }
\nonumber \\
{X^I}_{\bf k, l} &=&+ {\cal M}_{ {\bf k}, {\bf k}'  } \, \left( \,
M \cdot {X^{I {\rm T}} }_{  {\bf l}', {\bf k}' }
\cdot M^{-1} \, \right) \, {{\cal M}^{-1}}_{ \bf l', l} \quad .
\label{extendedcp}
\end{eqnarray}
where ${\cal M}_{\bf k,l}$ is a unitary matrix, whose choice should be
determined from consistency with T-duality. In our notation, the Wilson line 
along $2$-direction is suppressed, but its presence should be borne in mind 
throughout. In fact, this Wilson line breaks the U(2N) covering space gauge 
group into U(N)$\times$U(N). 

In the previous work~\cite{kimrey1}, we have classified all possible choices of
${\cal M}_{\bf k, l}$ from consistency with area-preserving diffeomorphism
gauge symmetry, and have found that ${\cal M}_{\bf k,l} = (-)^{\bf k 
\cdot e } \, \delta({\bf k} - \hat {\bf l})$, 
where $\hat{k} \equiv (-k_1, k_2)$.
This choice is unique~\footnote{The
other consistent choice is ${\cal M}_{\bf k, l}= (-)^{{\bf k} \cdot
{\bf e}_1} \,
\delta ({\bf k} - \hat {\bf l})$, but the final form of Chan-Paton condition
is unchanged.} and, as we will see momentarily, turns out to be\
the choice also consistent with T-duality of M(atrix) theory.
Using this result and the matrices $X^\mu({\bf k})$ introduced in section 2,
the Chan-Paton condition now reads
\begin{eqnarray}
A_0 ({\bf k}, t) &=&- \, (-)^{\bf k \cdot e} \, 
M \cdot {A_0}^{\rm T} (-\hat {\bf k}, t) \cdot M^{-1} \\
X^1 ({\bf k}, t)  &=&- \, (-)^{\bf k \cdot e} \, 
M \cdot X^{1{\rm T}} (-\hat{\bf k}, t) \cdot M^{-1} \\
X^2 ({\bf k}, t) &=& + \, (-)^{ \bf k \cdot e } \, 
M \cdot X^{2{\rm T}} (-\hat {\bf k}, t) \cdot M^{-1} \\
X^I ({\bf k}, t) &=&+ \, (-)^{\bf k \cdot e} \, 
M \cdot X^{I{\rm T}} (-\hat {\bf k}, t) \cdot M^{-1}.
\end{eqnarray}
The relative minus sign in the argument of right hand side 
is reflects the fact that cell indices in Eq.~(\ref{extendedcp})
are transposed so that the condition becomes compatible with single cell
condition Eq.~(\ref{originalkbmatrix}). 
Fourier transforming $X^\mu ({\bf k})$ to (2+1)-dimensional fields,  
\begin{eqnarray}
A_\alpha ({\bf y}, t) &=&\sum_{{\bf k}}A_\alpha ({\bf k}, t) \,
\exp \Big[ i  {\bf k} \cdot \widetilde{G} \cdot {\bf y} \Big] 
\\
Y^I({\bf y}, t) &=&\sum_{{\bf k}}
\, {Y^I} ( {\bf k}, t) \,
\exp \big[ i  {\bf k} \cdot \widetilde{G} \cdot {\bf y}   \Big] \quad ,
\end{eqnarray}
where $\alpha =0,1,2$, the above Chan-Paton conditions turn into 
\begin{eqnarray}
A_0({\bf y}, t) &=& - \, M \cdot {A_0}^{\rm T}
( - \hat {\bf y} + \pi \widetilde {G} \cdot {\bf e}, t) \cdot M^{-1} 
\nonumber\\
A_1({\bf y}, t) &=&- \, M \cdot {A_1}^{\rm T}(
  - \hat {\bf y} + \pi \widetilde {G} \cdot {\bf e} , t ) \cdot M^{-1} 
\nonumber \\
A_2({\bf y}, t) &=&+ \, M \cdot {A_2}^{\rm T}(
  - \hat {\bf y} + \pi \widetilde {G} \cdot {\bf e}, t ) \cdot M^{-1} 
\nonumber \\
Y^I({\bf y}, t) &=&+ \, M \cdot Y^{I{\rm T}}
(  - \hat {\bf y} + \pi \widetilde {G} \cdot {\bf e}, t ) \cdot M^{-1} 
\nonumber \\
\psi_A({\bf y}, t) &=& \! \Gamma_\perp  M
\cdot {\psi_A}^{\rm T} ( -\hat {\bf y} + \pi \widetilde {G} \cdot {\bf e}, t) 
\cdot M^{-1} \quad,
\label{gencpcond}
\end{eqnarray}
where again $\hat{\bf y} \equiv (-y_1, y_2),  \quad 
{\bf e} \equiv ({\bf e_1, e_2}), \quad
\Gamma_\perp = \gamma^0 \gamma^1, \quad  \left(\Gamma_\perp\right)^2  = 1$, 
and $\pm$ signs  are for two different choices of $M$.
With covering space Chan-Paton condition imposed, the M(atrix) 
theory turns into (2+1)-dimensional gauge theory whose covering space gauge 
group U(2N) is broken to U(N)$\times$U(N) by the presence of $A_2$ 
Wilson line.
Furthermore, the parameter space on which the gauge theory lives is
{\sl dual Klein bottle} $\widetilde{\bf K}_2$ since, 
according to  Eq.~(\ref{gencpcond}), 
the parameter space is a quotient of $\widetilde{\bf T}_2$ by
{\sl dual} symmetry group:
\be
\widetilde{\Gamma}_{\bf K} \quad : \quad
{\bf y} \rightarrow - \hat{\bf y} + \pi \widetilde{G} \cdot {\bf e},
\ee
viz. $\widetilde{\bf K}_2 = \widetilde{\bf T}^2 /
\widetilde{\Gamma}_{\bf K}$ of volume $\left(2 \pi \widetilde{R}_1 \right)
\times \left(\pi \widetilde{R}_2 \right)$. Note that in $\widetilde{\bf K}_2$
the coordinate that parity operation acts is interchanged compared to the
${\bf K}_2$.
We conclude that M(atrix) theory description of M-theory compactified
on Klein bottle ${\bf K}_2$ is defined by (2+1)-dimensional ${\cal N}
= 8$ supersymmetric U(N)$\times$U(N) gauge theory living on dual Klein
bottle $\widetilde{\bf K}_2$.

In fact, the Chan-Paton condition Eq.~(\ref{gencpcond}) can be seen to be 
consistent with T-duality of M(atrix) theory.
Recall that T-duality of M(atrix) theory is defined in terms of T-duality
of D0-parton themselves.  Thus, consider Type IIA string compactified
on Klein bottle ${\bf K}_2 = {\bf T}_2/\Gamma_{\bf K}$. 
The action of $\Gamma_{\bf K}$ is a product
of $({\cal P}_1 \Omega) \cdot {\cal S}$, where ${\cal P}_{1,2}$ are parity
inversion along $1,2$ directions, $\Omega$ worldsheet parity inversion and
${\cal S}$ half-period shift along both directions. 
Under T-duality along $1$-direction, the theory turns into Type IIB string 
compactified on $\left(\widetilde{\bf S}_1 \times {\bf S}_1\right)
/\left( {\cal S} \cdot \Omega \right)$.
Under another T-duality along $2$ direction, the worldsheet parity $\Omega$
is mapped into ${\cal P}_2 \cdot \Omega$, and the theory is turned into 
$\widetilde{\rm IIA}$ string compactified on $\widetilde{\bf T}_2/
\left({\cal P}_2 \cdot \Omega \cdot {\cal S} \right)$. The 
compactification is again on 
Klein bottle but with inverted volume and parity transformation direction 
compared to the starting Klein bottle ${\bf K}_2$. 
In fact, it is precisely the dual Klein bottle $\widetilde{\bf K}_2$ 
we have identified just above. The D0-partons are now T-dualized into 
D2-branes wrapped around the dual Klein bottle $\widetilde{\bf K}_2$ 
. The large-N limit of (2+1)-dimensional world-volume gauge theory of 
$\widetilde{\rm IIA}$ D2-branes on $\widetilde {\bf K}_2$
reduces to the Klein bottle M(atrix) theory we have deduced from 
the first-principle Chan-Paton conditions.

Having now understood the Chan-Paton conditions systematically,
we can make a short-cut derivation of the Klein bottle M(atrix) theory. 
Begin with (2+1)-dimensional ${\cal N} = 16$ supersymmetric $U(2N)$ gauge 
theory on $\widetilde{\bf T}_2$, relevant for M(atrix) theory on ${\bf T}_2$.
Consider the following set of transformations to the gauge theory. The first
is orientation reversal or, equivalently, complex conjugation
\bee
\Omega \hskip1cm : \hskip1cm 
A_0 ({\bf y}, t) & \rightarrow & \Omega \cdot A_0({\bf y}, t) \cdot
\Omega^{-1} = M \cdot {A_0}^{\rm T} ({\bf y}, t) \cdot M^{-1} \nonumber\\
A_1 ({\bf y}, t) & \rightarrow & \Omega \cdot A_1 ({\bf y}, t) \cdot
\Omega^{-1} = M \cdot {A_1}^{\rm T} ({\bf y}, t) \cdot M^{-1} \nonumber\\
A_2 ({\bf y}, t) & \rightarrow & \Omega \cdot A_2 ({\bf y}, t) \cdot
\Omega^{-1} = M \cdot {A_2}^{\rm T} ({\bf y}, t) \cdot M^{-1} \nonumber\\
Y^I ({\bf y}, t) & \rightarrow & \Omega \cdot Y^I({\bf y}, t) \cdot
\Omega^{-1} = M \cdot Y^{I{\rm T}} ({\bf y}, t) \cdot M^{_1} \nonumber\\
\psi_A ({\bf y}, t) &\rightarrow& \Omega \cdot {\psi_A}
({\bf y}, t) \cdot \Omega^{-1} = M \cdot {\psi_A}^{\rm T}({\bf y}, t) 
\cdot M^{-1} \quad ,
\label{orientation}
\eee
where $M$ is an arbitrary matrix subject to Hermiticity condition
$M^{-1} \cdot M^{\rm T} = \pm {\bf I}$,  
\bee
{\cal S} \hskip1cm : \hskip1cm
A_0 ({\bf y}, t) & \rightarrow & 
{\cal S} \cdot A_0 ({\cal S} \cdot {\bf y}, t) \cdot {\cal S}^{-1} = 
A_0 ({\bf y} + \pi \widetilde {G} \cdot {\bf e}, t) \nonumber\\
A_1 ({\bf y}, t) & \rightarrow & {\cal S} \cdot A_1 ({\cal S} \cdot
{\bf y}, t) \cdot {\cal S}^{-1} = 
A_1 ({\bf y} + \pi \widetilde {G} \cdot {\bf e}, t) + 
\left(\pi \widetilde{G} \right) \cdot {\bf e}_1 \nonumber\\
A_2 ({\bf y}, t) & \rightarrow & 
{\cal S} \cdot A_2 ({\cal S} \cdot {\bf y}, t) \cdot {\cal S}^{-1} =
A_2 ({\bf y} + \pi \widetilde {G} \cdot {\bf e}, t) +
\left( \pi \widetilde{G} \right) \cdot {\bf e}_2 \nonumber\\
Y^I ({\bf y}, t) & \rightarrow & {\cal S} \cdot Y^I ({\cal S} \cdot {\bf y}, t)
\cdot {\cal S}^{-1} = Y^I ({\bf y} + \pi \widetilde {G} \cdot 
{\bf e}, t) 
\nonumber\\
\psi_{A} ({\bf y}, t) &\rightarrow& {\cal S} \cdot \psi_{A}(
{\cal S} \cdot {\bf y}, t) \cdot {\cal S}^{-1} = \psi_{A} 
({\bf y} + \pi \widetilde{G} \cdot {\bf e}, t),
\label{shift}
\eee
the half-period shift transformations along both directions of $\widetilde
{T}_2$ and gauge connections, and
\bee
{\cal P} \hskip1cm : \hskip1cm
A_0 ({\bf y}, t) & \rightarrow & {\cal P} \cdot A_0 ({\cal P} \cdot {\bf y},
t) \cdot {\cal P}^{-1} = + \, A_0 (- \hat {\bf y}, t) \nonumber\\
A_1 ({\bf y}, t) & \rightarrow & {\cal P} \cdot A_1 ({\cal P} \cdot {\bf y},
t) \cdot {\cal P}^{-1} = + \, A_1 (- \hat {\bf y}, t) \nonumber\\
A_2 ({\bf y}, t) & \rightarrow & {\cal P} \cdot A_2 ({\cal P} \cdot {\bf y}, 
t) \cdot {\cal P}^{-1} = - \, A_2 (- \hat {\bf y}, t) \nonumber\\
Y^I ({\bf y}, t) & \rightarrow & {\cal P} \cdot Y^I ({\cal P} \cdot {\bf y},
t) \cdot {\cal P}^{-1} = - \, Y^I (- \hat {\bf y}, t) \nonumber\\
\psi_{A} ({\bf y}, t) &\rightarrow& {\cal P} \cdot \psi_{A}
({\cal P} \cdot {\bf y}, t) \cdot {\cal P}^{-1} 
= \Gamma_\perp \psi_{A} (- \hat{\bf y}, t),
\label{parity}
\eee
the parity transformation in (2+1) dimensions, 
where we have used the fact that $Y^I$'s are pseudo-scalars.

It is straightforward to check that the starting (2+1) dimensional gauge 
theory is invariant under a simultaneous tranformation $\Omega \cdot {\cal S} 
\cdot {\cal P}$ of Eqs.~(\ref{orientation} - \ref{parity}). 
In fact, the $\integer_2$ action $\Omega \cdot {\cal S} \cdot
{\cal P}$ is exactly identical to the covering space Chan-Paton condition Eq.
(\ref{gencpcond}). 
Hence, if we mod out the theory by this $\integer_2$ symmetry group,
we obtain precisely the Klein bottle M(atrix) theory defined on dual Klein
bottle $\widetilde{\bf K}_2$. 
Allowed choices of $M$ in Eq.~(\ref{orientation})
are precisely the ones $M = {\bf I} \otimes \sigma^1$ and 
${\bf I} \otimes \sigma^2$ permitted in Eq.~(\ref{gencpcond}).
Moreover, it is to be noted that the action of ${\cal S}$ in Eq.~(\ref{shift})
is accompanied by turning on constant $A_{1,2}$ gauge field backgrounds.
This background then breaks the starting U(2N) gauge group down to  
U(N)$\times$U(N) and no further gauge symmetry enhancement is permitted. 
Intuitively this can be understood from the fact that the
defining $N$ D0-partons and their mirror partons never come close due to 
the fact that the $\integer_2$ involution is free.

To summarize, in M(atrix) theory, nine-dimensional M-theory compactified 
on Klein bottle is described by (2+1)-dimensional ${\cal N} = 8$ 
supersymmetric U(N)$\times$U(N) gauge theory living on {\sl dual Klein bottle} 
parameter space.

\subsection{BPS Branes and Massless Spacetime Spectra}
In M(atrix) theory various BPS branes arise as composite bound-states of 
D0-partons. By identifying BPS branes that are consistent with the orbifold 
projection $\Gamma$, as proposed and utilized in other orbifolds~\cite{kimrey1}, it is possible to extract massless spacetime spectrum.
We adopt the same strategy and deduce $d=9$ spacetime spectrum now.
To do so, we utilize two possible types of D0-parton bound-states.
Two types of D0-parton bound-states are threshold bound-states representing
M-theory graviton and Landau-level orbiting bound-states representing
electric BPS states coupled to M-theory three-form potential.
Threshold bound-states represent graviton. 
Projecting out odd components of the bound-state under $\Gamma_{\bf K}$
, hence, $\widetilde{\Gamma}_{\bf K}$, one finds that the threshold
bound-state of D0-parton give rise only to dimensionally reduced gravitons. 
The Landau-orbiting bound-state corresponds in M(atrix) gauge theory to   
a quantum of magnetic flux ${\bf B}$, represented by half-integer
valued first Chern class
\be
 {\rm Tr} \int_{\widetilde{\bf K}_2} {d^2 {\bf y} \over 2 \pi} \, {\bf B}
= {m \over 2}
\quad, \quad \quad m \in \integer \quad ,
\ee
and gives rise to membrane configuration wrapped around the compactified 
Klein bottle. As is easily verified, this configuration is invariant under 
the $\Gamma_{\bf K}$ action. In the limit the area of M\"obius strip 
vanishes, infinite tower of wrapping states become massless densely and
the Aspinwall-Schwarz dimension (9-th) dimension opens up.   

The second is a quantum of electric flux ${\bf E}_2$ on $\widetilde{\bf K}_2$
\be
 {\rm Tr} \int_0^{\pi \widetilde{R}_2} d {\bf y} \cdot  {\bf E}_2
= {\gym \widetilde{R}_2 \over 4 \widetilde{R}_1} n \quad , \quad \quad
n \in \integer \quad.
\ee
The third is a photon propagating around the $x^1$ direction.
Each of these three configurations in gauge theory represent 
M-theory BPS states that couple minimally to spacetime gauge fields.
Thus, spacetime massless spectrum can be inferred from BPS spectrum of
M(atrix) gauge theory. 

Membrane located in ${\bf R}_9$ is described by
\be
Y^I  
= \left( \begin{array}{cc} P & 0 \\ 0 & + P^{\rm T} \end{array} \right)
\hskip1cm
Y^J 
= \left( \begin{array}{cc} Q & 0 \\ 0 & + Q^{\rm T} \end{array} \right)
\ee
so that
\be
{\cal Z}_2 = [Y^I, Y^J]
= {1 \over N} \left( \begin{array}{cc}
{\bf I} & 0 \\ 0 & - {\bf I} \end{array} \right)
\ee
Total BPS charge vanishes and is not compatible with $\Gamma$ projection.
Thus, there is no membrane propagating in ${\bf R}_9$ and we conclude 
that there is no massless three-form potential in the spacetime spectrum.

Membrane partially wrapped on ${\bf K}_2$ is given by 
\be
A_1 = \left( \begin{array}{cc} P & 0 \\ 0 & - P^{\rm T} \end{array} \right)
\hskip1cm
Y^I = \left( \begin{array}{cc} Q & 0 \\ 0 & + Q^{\rm T} \end{array} \right)
\ee 
carries
nontrivial BPS charge. Propagation of the state compatible with 
$\Gamma$ projection is in ${\bf R}_9$. Therefore, we conclude that there is
two-form potential in the spacetime spectrum. 
The commutator is $\integer_2$ even, hence, gives rise to a consistent
configuration.
On the other hand, membrane partially wrapped around the other direction
of ${\bf K}_2$ carries no BPS charge, hence, does not give rise to second
two-form potential in the spectrum.
Also, electric field excitation gives rise to BPS particle state, which 
couples minimally to the $B_{9 \mu}$ component.
Altogether, we have found that the spacetime spectrum includes graviton,
single two-form tensor potential and dilaton. This is precisely the spectrum
of Type IIB string upon compactification on ${\bf S}_1/\Gamma_2$. 

The parameter space of the M(atrix) gauge theory 
is Klein bottle, hence, do not have any orbifold fixed boundaries.
Since the gauge theory in that case has no room for potential gauge 
nor supersymmetry anomalies, we conclude that there is no twisted sector
in the M(atrix) theory. This in turn implies that there is no {\sl charged} 
states, hence, no gauge symmetry group in spacetime spectrum.

\section{M(atrix) Theory on M\"obius strip}
Next, consider another possible nine-dimensional M-theory compactification on
a M\"obius strip ${\bf M}_2$. 
M\"obius strip ${\bf M}_2$ of area $\left( w \pi R_1 \right) \times \left(
2 \pi R_2 \right) / 2 $ may be obtained as a quotient of torus ${\bf T}_2$
of area $\left(2 \pi R_1 \right) \times \left( 2 \pi R_2 \right)$ by 
symmetry group
\be
\Gamma_{\bf M} \quad : \quad {\bf x} \rightarrow
\widetilde{\bf x}, \ee
where $\widetilde{\bf x} \equiv (x^2, x^1)$, and is encompassed by fundamental domain
$0 \le x^1 \le \left( R_1 / R_2 \right) x^2, \quad 0 \le x^2 \le 2 \pi R_2$.
.
After rotation of the coordinates by $\pi/4$, the M\"obius strip constructed 
as above is equivalent to 
$\left( {\bf T}_2 / {\cal P} \cdot \Omega \right) / {\cal S} $
, viz. $\integer_2$ quotient of cylinder.
To define the corresponding M\"obius M(atrix) theory, we put $N$ D0-branes
on ${\bf M}_2$ and study their dynamics on covering space ${\bf T}_2$. On 
${\bf T}_2$, this amounts to putting $2N$ D0-partons. They form a single 
$\integer_2$ orbit of the symmetry group $\Gamma_{\bf M}$ on ${\bf T}_2$.  
M-theory
compactification on M\"obius strip has been proposed as a dual to strong 
coupling limit of CHL string~\cite{chaupol, dabpark}.
In this section, we study the corresponding M(atrix) theory and show that
indeed the CHL string spectrum as well as Wilson line moduli space follow
from the theory. 

\subsection{Chan-Paton Condition}
Again we describe M\"obius M(atrix) theory in terms of dynamics among
$2N$ D0-partons on a single fundamental cell of covering space ${\bf T}_2$.
The Chan-Paton conditions to D0-partons on the single fundamental cell 
are given by
\begin{eqnarray}
A_0 &=& \, - \, M \cdot {A_0}^{\rm T} \cdot M^{-1} \nonumber \\
X^1 &=& \, + \, M \cdot X^{2{\rm T}} \cdot M^{-1} \nonumber \\
X^2 &=& \, + \, M \cdot X^{1{\rm T}} \cdot M^{-1} \nonumber \\
X^I &=& \, + \, M \cdot X^{I{\rm T}} \cdot M^{-1} \hskip2.81cm
(I=3, \cdots, 9),                          \nonumber \\
\psi_A &=&  \Gamma_\perp M \cdot {\psi_A}^{\rm T} \cdot M^{-1},
\hskip2.5cm
\Gamma_\perp \equiv {i \over 2} (\gamma^1 - \gamma^2),
\end{eqnarray}
where the spinor projection operator $\left( \Gamma_\perp \right)^2
= 1$. We now extend the Chan-Paton
condition over the image ${\bf T}_2$ cells in the covering space ${\bf R}_2$.
Following exactly the same procedure as Klein bottle compactification case, 
we find
\bee
A_{0 \bf k, m} &=& \, - \, {\cal M}_{\bf k, k'} \, (M
\cdot {{A_0}^{\rm T}}_{\bf m', k'} \cdot
M^{-1}) \, {{\cal M}^{-1}}_{\bf m', m} \nonumber \\
{X^1}_{ \bf k, m} &=& \, +\, {\cal M}_{\bf k, k'} \,( M
\cdot {X^{2\rm T}}_{\bf m', k'} \cdot
M^{-1} ) \, {{\cal M}^{-1}}_{\bf m', m} \nonumber \\
{X^2}_{ \bf k, m} &=& \, +\,  {\cal M}_{\bf k, k'} \,( M
\cdot {X^{1 \rm T}}_{\bf m', k'} \cdot
M^{-1}) \, {{\cal M}^{-1}}_{\bf m', m} \nonumber \\
{X^I}_{ \bf k, m} &=& \, + \, {\cal M}_{\bf k, k'} \, ( M
\cdot {X^{I\rm T}}_{\bf m', k'} \cdot
M^{-1} ) \, {{\cal M}^{-1}}_{\bf m', m} ,
\nonumber \\
\psi_{A \bf k, m} &=& \Gamma_\perp {\cal M}_{\bf k, k'} ( M \cdot
{{\psi_A}^{\rm T}}_{\bf m', k'} \cdot M^{-1} ) {{\cal M}^{-1}}_{\bf m', m}.
\eee

In the previous work~\cite{kimrey1}, we have also identified possible 
choices of ${\cal M}_{\bf k,l}$ for M\"obius strip from the 
consistency of area-preserving diffeomorphism gauge symmetry. 
It was found there that a unique choice is 
${\cal M}_{\bf k, m} = \delta({\bf k} - \widetilde {\bf m})$~\footnote{
Alternative possible choice is ${\cal M}_{\bf k, m} = (-)^{{\bf k} \cdot 
({\bf e}_1 + {\bf e}_2)} \, \delta({\bf k} - \widetilde {\bf m})$,
but again, this choice results in equivalent covering space Chan-Paton 
conditions.} where $\widetilde{\bf m} \equiv (m_2, m_1)$. 
The Chan-Paton conditions on the covering space is found to be 
\begin{eqnarray}
A_0 ({\bf k}, t) &=& - \, M \cdot {A_0}^{\rm T}  (-\widetilde{\bf k}, t) \cdot M^{-1} \nonumber \\
X^1 ({\bf k}, t) &=& + \, M \cdot X^{2{\rm T}} (-\widetilde{\bf k}, t) \cdot M^{-1} \nonumber \\
X^2 ({\bf k}, t) &=& + \, M \cdot X^{1{\rm T}} (-\widetilde{\bf k}, t) \cdot M^{-1} \nonumber \\
X^I ({\bf k}, t) &=& + \, M \cdot X^{I{\rm T}} (-\widetilde{\bf k}, t) \cdot M^{-1}
\nonumber \\
\psi_{A} ({\bf k}, t) 
&=& \Gamma_\perp M \cdot {\psi_{A}}^{\rm T} (- \widetilde{\bf k}, t)
\cdot M^{-1} \quad .
\end{eqnarray}
Again, the relative minus sign in the argument on the right hand side is a 
direct reflection of the transposition in the covering space Chan-Paton condition.  In terms of Fourier transformed fields , the Chan-Paton conditions 
become
\begin{eqnarray}
A_0({\bf y}, t) &=& - \, M \cdot {A_0}^{\rm T}(-\widetilde{\bf y}, t)
\cdot M^{-1}  \nonumber \\
A_1({\bf y}, t) &=& + \, M \cdot {A_2}^{{\rm T}}(-\widetilde{\bf y}, t)
\cdot M^{-1}  \nonumber \\
A_2({\bf y}, t) &=& + \, M \cdot {A_1}^{{\rm T}}(-\widetilde{\bf y}, t)
\cdot M^{-1} \nonumber \\
Y^I({\bf y}, t) &=& + \, M \cdot Y^{I{\rm T}}(-\widetilde{\bf y}, t)
\cdot M^{-1} \nonumber \\
\psi_{A} ({\bf y}, t) &=& \Gamma_\perp M \cdot {\psi_{A}}^{\rm T} 
(-\widetilde{\bf y}, t) \cdot M^{-1} \quad.
\label{generalmbcond}
\end{eqnarray}
With these covering space Chan-Paton conditions imposed, the M(atrix) 
theory turns into (2+1)-dimensional gauge theory on {\sl dual M\"obius 
strip} $\widetilde{\bf M}_2$, since according to Eq.~(\ref{generalmbcond}) ,
the parameter space is a quotient of $\widetilde{\bf T}_2$ by {\sl
dual} symmetry group:
\be
\widetilde{\Gamma}_{\bf M} \quad : \quad
{\bf y} \rightarrow - \widetilde{\bf y}.
\ee
Hence, $\widetilde{\bf M}_2 = \widetilde{\bf T}_2/\widetilde{\Gamma}_{\bf M}$.
Note that the boundary direction of original and dual M\"obius strips
is exchanged each other.

The above covering space Chan-Paton condition is again consistent with 
T-duality of M(atrix) theory. To exhibit this, recall that the M\"obius 
strip ${\bf M}_2$ is obtained as a quotient of cylinder ${\bf C}_2 
= \left({\bf S}_1 / {\cal P}_1 \cdot \Omega \right) \times {\bf S}_1$ by 
$\integer_2$ symmetry group ${\cal S}$ that shifts both coordinates of
covering space ${\bf T}_2$ by half periods. Thus, D0-parton dynamics on
${\bf M}_2$ may be understood in terms of ${\cal S}$ projection of
Type IA / heterotic string theories. Under T-duality along $1$-direction,
the theory is mapped into Type I string on $\left(\widetilde{\bf S}_1 
\times {\bf S}_1 \right)/{\cal S}$. The D0-partons are mapped into
D1-strings of Type I string theory. Upon another T-duality along 2-direction,
the worldsheet parity inversion $\Omega$ is mapped into $\Omega \cdot 
{\cal P}_2$. The resulting theory is Type $\widetilde{\rm IA}$ string 
compactified on $\left( \widetilde{\bf S}_1 \times 
(\widetilde{\bf S}_1/{\cal P}_2 \cdot \Omega) \right) / {\cal S}$. 
The resulting compactification space is again a M\"obius strip but now 
with inverted volume and exchanged parity tranformation direction.  
 
We can proceed directly from (2+1)-dimensional ${\cal N}=16$ supersymmetric
U(4N) gauge theory on $\widetilde{\bf T}_2$ and obtain a short-cut derivation
of the above Chan-Paton conditions.
It is then straightforward to recognize that the starting gauge theory is 
invariant under a combined action of $\Omega \cdot {\cal P}_-$, where
$\Omega$ is the orientation reversal or complex conjugation and 
${\cal P}_-$ is parity inversion along the diagonal:
\bee
{\cal P}_- \hskip0.4cm : \hskip0.4cm
A_0 (y^\pm, t) &\rightarrow& 
{\cal P} \cdot A_0
({\cal P} \cdot y^\pm, t) \cdot {\cal P}^{-1}
= + \, A_0 (\mp y^\pm, t) \\
A_-(y^\pm, t) &\rightarrow& 
{\cal P} \cdot A_- ({\cal P} \cdot y^\pm, t) \cdot {\cal P}^{-1}
= + \, A_-(\mp y^\pm, t) \\
A_+(y^\pm, t) &\rightarrow&
{\cal P} \cdot A_+ ({\cal P} \cdot y^\pm, t) \cdot {\cal P}^{-1}
= - \, A_+(\mp y^\pm, t) \\
Y^I(y^\pm, t) &\rightarrow& 
{\cal P} \cdot Y^I
({\cal P} \cdot y^\pm, t) \cdot {\cal P}^{-1} =
- \, Y^I (\mp y^\pm, t).
\\
\psi_A(y^\pm, t) &\rightarrow&
{\cal P} \cdot \psi_A ({\cal P} \cdot y^\pm, t) \cdot {\cal P}^{-1}
=
i \gamma^- \psi_A(\mp y^\pm, t).
\eee
where $y^\pm \equiv y^1 \pm y^2, A_\pm = A_1 \pm A_2$
and $\gamma^\pm = {1 \over \sqrt{2}} (\gamma^1 \pm \gamma^2)$.
Modding out the defining gauge theory by $\Omega \cdot {\cal P}_-$, we obtain 
precisely the same conditions as the covering space Chan-Paton conditions
Eq.~(\ref{generalmbcond}).

To summarize, M\"obius M(atrix) theory is described by
(2+1)-dimensional ${\cal N} = 8$ supersymmetric U(N)$\times$U(N) 
gauge theory on the {\sl dual Klein bottle} $\widetilde{{\bf K}_2}$:
\bee
L_{\rm untwisted}
= -{1 \over g^2_{\rm YM}} \int_{\widetilde {\bf M}_2}
d^2 {\bf y} \,\, {\rm Tr} \Big[
&& \hskip-0.4cm 
F_{\alpha \beta} F^{\alpha \beta} + 2 D_\alpha Y_i D^\alpha Y^i
- [Y_i, Y_j] [Y^i, Y^j]
\nonumber \\
&& - 2i {\overline \psi}_A \gamma^\alpha D_\alpha \psi_A
+ 2i {\overline \psi}_A \gamma^i_{AB} [Y^i, \psi_B] \, \Big]\,\,.
\eee
Note that an overall factor of two has been inserted since we write the
action only on the fundamental domain of the M\"obius strip.
This (2+1)-dimensional bulk gauge theory will be called as {\sl untwisted
sector} of CHL M(atrix) theory. 

\subsection{Twisted Sector and Gauge Symmetry}
The $\integer_2$ involution has a orientifold fixed line $x = y$. 
The orientifold plane carries - 8 units of D8-brane charge as can be
seen, for example, from the D0-parton scattering off the orbifold fixed
circle $y^+ = 0$. In fact, at the orbifold fixed circle, from the covering
space Chan-Paton condition, the action of $\Omega \cdot {\cal P}_-$ imposes 
boundary conditions on the fields:
\bee
(A_0, A_-, \psi_{2A}) (y^-) &\hskip0.5cm& {\tt antisymmetric}
\nonumber \\
\partial_+ (A_0, A_-, \psi_{2A}) (y^-) &\hskip0.5cm& {\tt symmetric}
\nonumber \\
(A_+, Y^i, \psi_{1A})(y^-) &\hskip0.5cm& {\tt symmetric}
\nonumber \\
\partial_+ (A_+, Y^i, \psi_{1A})(y^-) &\hskip0.5cm& 
{\tt antisymmetric}
\eee

These boundary conditions modify the M(atrix) gauge theory in several ways.
First, they break half the supersymmetry. The dynamical supersymmetry 
parameters $\epsilon_A$ that appear in Eq.~(\ref{dynamicsusy}) must be taken to be invariant 
under the $\integer_2$ projection onto M\"obius strip, 
$\epsilon_A = i \gamma^- \epsilon_A$. This results in ${\cal N}= 8$ 
supersymmetry in 2+1 dimensions. 
Simultaneously, to respect the boundary conditions on the fermions, the 
kinematical supersymmetry parameters $\eta_A$ should satisfy $\eta_A = - i 
\gamma^- \eta_A$. The relative sign difference between the dyanmical and
the kinematical supersymmetry projections reflects the fact that $\epsilon_A$
and $\eta_A$ belong to inequivalent representations of the (2+1)-dimensional
Clifford algebra.

At the orbifold fixed boundary, the gauge transformation $A_\alpha
\rightarrow U ( - i \partial_\alpha + A_\alpha ) U^{-1}$ respects the 
boundary conditions only if $U^{\rm T} \cdot U (y^+ = 0, y^-)
= \pm \identity$, viz. only gauge transformations in an O(2N) subgroup of
U(2N) are permitted.
Under the boundary O(N) gauge group, the fields { $A_0, A_1, \psi_{2A}$ }
transform as the adjoint representation, while { $A_2, Y^i, \psi_{1A}$
} transform as the symmetric representation. The fermions $\psi_A$ have
normalizable modes which are independent of $y^+$. These modes behave
as (1+1)-dimensional spinors. From (2+1)-dimensional Dirac equation, it
is straightforward to see that $\psi_{1A}$ and $\psi_{2A}$ are in opposite 
chirality in the (1+1)-dimensional world. These modes therefore generate O(2N)
gauge anomaly $[8 I_2 ({\tt adj.}) - 8 I_2 ({\tt symm.}) ]/2 = 
- 16 I_2 ({\tt fund.})$, where the factor of 1/2 comes from the 
$\Gamma_2$ involution in defining the M\"obius strip from the cylinder.

This gauge anomaly is cancelled by introducing a twisted sector consisting
of sixteen left-moving Majorana-Weyl fermions $\chi(y^-)$ in the 
fundamental representation of O(2N). 
Much in parallel to the case of heterotic M(atrix) theory on cylinder, 
it is expected that this is also the choice that cancels potential 
supersymmetry anomaly. 
In order to cancel the anomaly {\sl locally}, the fermions should be localized
at the boundary of M\"obius strip. 

The twisted sector of M\"obius M(atrix) theory involves Majorana-Weyl 
fermions $\chi_M$. They are left-moving, hence, are (0,8) supersymmetry
singlets. They also transform in the fundamental representations of the
boundary M(atrix) gauge group O(2N) $\subset$ U(2N) as well as in the 
global `flavor' symmetry group SO(16) associated with the eith D8-branes 
and their images present at the boundary. 
Their Lagrangian is 
given by
\be
L_{\rm twisted} = \oint_{\partial {\bf M}_2} d y^- \,
i \chi_M \Big( \delta_{MN} (D_0 + D_-) +i (B_0 + B_-)^{MN} \Big)_{y^+ = 
0} \chi_N.
\ee
It is tacitly assumed that (1+1)-dimensional boundary dynamics of the 
gauge fields $A_\pm $ is part of the untwisted sector Lagrangian.
We also have included couplings to the background field $B_{MN}$ which 
are in the adjoint representation of SO(16).
This is in fact the spacetime gauge field of strongly coupled CHL 
heterotic string, hence, turning on $B_{MN}$ corresponds to turning on 
Wilson line moduli.

\subsection{Chern-Simons Coupling and Wilson Line Moduli}
An interesting and important question is to understand the Narain moduli
space of strongly coupled CHL string in terms of M(atrix) theory.
The reason why this is so is because the moduli spaces of CHL string contain
points of not only simply-laced but also non-simply-laced enhanced symmetry,
as well as higher Kac-Moody level realization of gauge symmetry.
In the strongly coupled limit, how are these features realized?
For example, in D=4 compactifications, it has been known that Sp(2n) and
SO(2n+1) are interchanged under electric-magnetic duality. 

In this section, as a first step toward a complete understanding of these
issues, we address how the Wilson line moduli space is realized in M(atrix)
theory.

As in $E_8 \times E_8$ heterotic M(atrix) theory on ${\bf S}_1$, the Wilson
line is realized in terms of the positions of twisted sector Majorana-Weyl
spinors on the dual M\"obius strip. We have argued that these spinors are
$y^+$-independent normalizable zero modes on a circle parallel to the
boundary of M\"obius strip. Geometric moduli of moving them away from the 
boundary and splitting among themselves then correspond to realization of 
Wilson line moduli of strongly coupled CHL string.
In order for this deformation to be compatible with {\sl local} cancellation
of supersymmetry and gauge anomalies, it is necessary to include Chern-Simons
term to the M(atrix) theory.
Turning on Chern-Simons term is actually more involved, since it is directly
a result of massive Type IIA supergravity background. Nevertheless, the 
procedure is essentially the same as that for heterotic M(atrix) theory. 
It is convenient to introduce new fields
\be
Z^i = \left(z(y^+)\right)^{1/3} Y^i
\ee
The
final M\"obius M(atrix) theory Lagrangian is then given by:
\bee
\label{MatrixAction}
L_{\rm Mobius} & = & - {1 \over \gymb} \int d^2 y \, 
{\rm Tr} \bigg\lbrace
z(y^+) F_{\alpha\beta} F^{\alpha\beta} + 2 z^{1/3}(y^+) 
D_\alpha Z_i D^\alpha Z^i - z^{-1/3}(y^+) [Z_i,Z_j][Z^i,Z^j] \nonumber \\
& & \quad \qquad \qquad \qquad - 2 i z^{1/3}(y^+)
\overline{\psi}_A \gamma^\alpha D_\alpha \psi_A - {d z^{1/3} \over
dy^+} \overline{\psi}_A \psi_A + 2 i \overline{\psi}_A \gamma^i_{AB}
[Z_i, \psi_B] \nonumber \\
& & \quad \qquad \qquad \qquad - {4 \over 3} {dz \over dy^+}
\epsilon^{\alpha \beta \gamma}
\bigl(A_\alpha \partial_\beta A_\gamma + i {2 \over 3} A_\alpha
A_\beta A_\gamma \bigr) \bigg\rbrace \\
& & + i \sum_{M,N=1}^{8} \oint d y^- \, 
\overline{\chi}_M \left( \delta^{MN}(D_0 + D_-) + i (B_0 + B_-)^{MN} 
\right) \chi_N
\,\,\,. \nonumber
\eee



\section{M(atrix) Theory of $d=6$ CHL Compactification}

So far, we have studied M(atrix) theory on the simplest, two-dimensional 
non-orientable manifolds.
In this section, we extend our study to higher dimensional Ricci flat 
non-orientable manifold.
Consider CHL heterotic string compactified toroidally down to six-dimensions.
In strong coupling limit, this theory is described as a quotient of M-theory 
compactified on $({\bf S}_1 / \integer_2) \times {\bf T}^4$ by the CHL projection group $\Gamma_2^{\rm CHL}$, which acts as half-period shifts 
:
\be
\Gamma^{\rm CHL}_2 \quad : \quad
{\bf x} \rightarrow {\bf x} + \left(\pi G \right) \cdot {\bf e},
\ee
of ${\bf x} = (x^1,x^2)$ coordinates along 
${\bf S}_1/\integer_2$ and one of the four coordinates of ${\bf T}_4$. 
Such half-period shift has previously been used for an extensive construction 
of string dual pairs~\cite{schwarzsen, vafawitten}. 
After rotating the coordinates by $\pi / 4$, it is straightforward to recognize 
that the simultaneous action of $\integer_2$ defining the cylinder ${\bf C}_2$
and $\Gamma_2^{\rm CHL}$ gives rise to the involution defining 
M\"obius strip ${\bf M}_2$ out of covering space torus ${\bf T}_2$. 
It is always implicit that the quotient is accompanied by $\Omega$ that flips
 the sign of the three-form potential $C_{MNP}$.  

What is the relevant M(atrix) theory description of this strong coupling 
CHL compactification? For ${\bf T}_5$ 
and ${\bf T}_5/ \integer_2$ compactifications, it has been suggested that 
the relevant M(atrix) theory is six-dimensional {\sl little} string theory 
with (2,0) or (1,0) chiral supersymmetries~\cite{seiberg}. 
Likewise, we expect that  
strong coupling CHL compactification on ${\bf M} \times {\bf T}_3$ is
described by a six-dimensional {\sl little} string theory with (1, 0)
chiral supersymmetry living on orbifold limit parameter space of 
$( {\bf S}_1 \otimes K3 ) / \integer_2$. Here, the $\integer_2$ involution 
of the paramter space acts simultaneously on ${\bf S}_1$ as a half-period 
shift and on $K3$ as an involution $\sigma$.
As a modest check of this conjecture, we show below that the low-energy
dynamics of the M(atrix) theory, viz. (1,0) {\sl little} string theory, 
is given by (5+1)-dimensional (1,0) supersymmetric U(N)$\times$U(N) 
gauge theory
living on {\sl dual} parameter space $( {\tilde {\bf S}}_1 \otimes \tilde 
{K3} ) / \integer_2$.
The low-energy effective description via supersymmetric gauge theory 
is obtained essentially by the same procedure as the two-dimensional
compactifications. 
For vanishingly small size of ${\bf M}_2 \otimes
{\bf T}_3$, we properly take into account of all possible winding string
configurations between D0-partons via Fourier transformation.
In terms of covering space description, 
\be
{\bf M}_2 \otimes {\bf T}_3 = \Gamma \backslash {\bf R}^5 / \integer^5 ,
\nonumber
\ee 
D0-parton dynamics on covering space is described by (1,0) supersymmetric 
U(N) gauge theory on {\sl dual} $\widetilde {\bf T}^5$ parameter space. 
It now remains to determine the action of an appropriate involution to 
this gauge theory. 

In the covering space ${\bf T}^5$,
low energy effective dynamics of M(atrix) theory is described by 
six-dimensional gauge theory on dual parameter space $\widetilde{\bf T}^5$ 
with ${\cal N} = 16$ supersymmetry and gauge group U(N). 
Let us denote the coordinates of parameter space $\widetilde{\bf T}^5$ as 
${\bf y} \equiv (y^1, y^2, \cdots, y^5)$. 
The orbifold under consideration is obtained by modding out by 
$\Gamma^{\rm CHL}_2  = {\cal P} \cdot \Omega$.
Acting on fields, they have the following action
\bee
\Omega \quad : \quad 
A_0 ({\bf y}, t) \rightarrow \Omega \cdot A_0 ({\bf y}, t) \cdot
\Omega^{-1} &=& - {A_0}^{\rm T} ({\bf y}, t)  
\nonumber \\
A_1 ({\bf y}, t) \rightarrow \Omega \cdot A_1 ({\bf y}, t) \cdot 
\Omega^{-1} &=& - {A_1}^{\rm T} ({\bf y}, t) 
\nonumber \\
A_2 ({\bf y}, t) \rightarrow \Omega \cdot A_2 ({\bf y}, t) \cdot
\Omega^{-1} &=& - {A_2}^{\rm T} ({\bf y}, t)
\nonumber \\
Y^I ({\bf y}, t) \rightarrow \Omega \cdot Y^I ({\bf y}, t) \cdot
\Omega^{-1} &=& - Y^{I{\rm T}} ({\bf y}, t) \quad , \eee 
where we have taken $M = {\bf I} \otimes \sigma^2$, 
\bee 
{\cal P} \quad : \quad 
A_0 ({\bf y}, t) \rightarrow {\cal P} \cdot A_0 ({\cal P} \cdot {\bf y}, t)
\cdot {\cal P}^{-1} &=& + {A_0} (-\hat {\bf y}, t) 
\nonumber \\
A_1 ({\bf y}, t) \rightarrow {\cal P} \cdot A_1 ({\cal P} \cdot {\bf y}, t)
\cdot {\cal P}^{-1} &=& + {A_1} (-\hat{\bf y}, t) 
\nonumber \\
A_2 ({\bf y}, t) \rightarrow {\cal P} \cdot A_2 ({\cal P} \cdot {\bf y}, t)
\cdot {\cal P}^{-1} &=& - {A_2} (-\hat{\bf y}, t) 
\nonumber \\
Y^I ({\bf y}, t) \rightarrow {\cal P} \cdot Y^I ({\cal P} \cdot {\bf y}, t)
\cdot {\cal P}^{-1} &=& - Y^{I} (-\hat{\bf y}, t) \quad ,
\eee
where $\hat{\bf y} \equiv {\cal P} \cdot {\bf y} = 
(y^1, - y^2, y^3, \cdots, y^5)$.
Finally, ${\cal S}$ is 
\bee
{\cal S} \quad : \quad
A_0 ({\bf y}, t) \rightarrow {\cal S} \cdot A_0 ({\cal S} 
\cdot {\bf y}, t) \cdot {\cal S}^{-1} &=& + {A_0} ({\bf y} + \pi
\widetilde{G} \cdot {\bf e}, t)
\nonumber \\
A_1 ({\bf y}, t) \rightarrow {\cal S} \cdot A_1 ({\cal S} \cdot {\bf y}, t)
\cdot {\cal S}^{-1} &=& + {A_2} ({\bf y} + \pi 
\widetilde{G} \cdot {\bf e}, t)         
\nonumber \\
A_2 ({\bf y}, t) \rightarrow {\cal S} \cdot A_2 ({\cal S} \cdot 
{\bf y}, t) \cdot {\cal S}^{-1} &=& - {A_1} ({\bf y} + \pi
\widetilde{G} \cdot {\bf e}, t)              
\nonumber \\
Y^I ({\bf y}, t) \rightarrow {\cal S} \cdot Y^I ({\cal S} \cdot {\bf y},t) 
\cdot {\cal S}^{-1} &=& + Y^{I} ({\bf y} + \pi
\widetilde{G} \cdot {\bf e}, t) \quad .
\nonumber \\
\eee
The combined action then acts on the covering space gauge theory fields as 
\bee
\Gamma^{\rm CHL}_2 \quad : \quad
A_0 ({\bf y}, t) &=& - {A_0}^{\rm T} (-\hat{\bf y} + \pi
\widetilde{G} \cdot {\bf e}, t)
\nonumber \\
A_1 ({\bf y}, t) &=& - {A_1}^{\rm T} (-\hat{\bf y} + \pi
\widetilde{G} \cdot {\bf e}, t)
\nonumber \\
A_2 ({\bf y}, t) &=& - {A_2}^{\rm T} (-\hat{\bf y} + \pi
\widetilde{G} \cdot {\bf e}, t)
\nonumber \\
Y^I ({\bf y}, t) &=& + Y^{I\rm T} (-\hat{\bf y} + \pi
\widetilde{G} \cdot {\bf e}, t) \quad .
\nonumber \\
\eee
It is then evident that the resulting gauge theory is defined
on a parameter space 
\be
{\left(\widetilde{\bf S}_1 \times (\widetilde{\bf T}^4/ \integer_2 )\right)
\over \widetilde{\Gamma}^{\rm CHL}_2}
\nonumber \\
\ee
where
\be
\widetilde{\Gamma}^{\rm CHL}_2 \quad : \quad
{\bf y} \rightarrow - \hat{\bf y} + \left( 
\pi \widetilde{G} \right) \cdot {\bf e}.
\nonumber
\ee
Thus the parameter space is $\widetilde{\Gamma}^{\rm CHL}_2$ quotient of
(orbifold limit of) $\widetilde{\bf S}_1 \times K3$. Note that the 
same parameter space has been identified as Type IIA compactification space
dual to the CHL heterotic string~\cite{schwarzsen}.
We conclude that strongly coupled CHL heterotic string compactified 
toroidally to six dimensions is described in M(atrix) theory by 
six-dimensional ${\cal N} = 8$ supersymmetric 
U(N)$\times$U(N) gauge theory living on $\widetilde{\bf S}_1 \times
K3$.

In fact, the resulting gauge theory can be understood via 
T-duality of M(atrix) theory as well.  We begin with the D0-partons living on 
$[({\bf S}_1 / \integer_2) \times {\bf T}_4 ]/\Gamma_2^{\rm CHL}$.
Rewriting the compactification as
\be
{\bf M}_2 \times {\bf T}_3 = [({\bf S}_1 / \integer_2) \times
{\bf T}_4 ]/\Gamma_2^{\rm CHL}
\ee
we now make T-duality tranformation successively.
Let us first T-dualize the ${\bf S}_1/\integer_2$ direction. This maps the
IIA D0-brane into IIB D1-string living on the space
\be
\left(
\tilde {\bf S}_1 \times {\bf T}_4 \right) / \Gamma_2^{\rm CHL}.
\ee
The D8-branes and $\Omega$8 orientifold plane that defines the twisted sector is mapped
to D9-branes and $\Omega_9$ plane.
Next, making T-duality transformations along the ${\bf T}_4$ directions,
the $\Omega_9$-plane and the D9-branes turn into $\Omega_5$-plane and
D9-branes into D5-branes. The IIA D0-partons are now mapped into D5-branes
wrapped around the compactified directions.
After the complete T-duality transformation, we have M(atrix) theory described
via Type IIB string compactified on
\be
{ \left( \widetilde {\bf S}_1 \times (\tilde {\bf T}_4 / \integer_2) \right)
\over \widetilde{\Gamma}_2^{\rm CHL} }
\ee
which is nothing but the dual parameter space identified just above.
The order-2 group $\Gamma_2^{\rm CHL}$ acts the same as before, viz.
half-period shift along dual circle direction and one of the dual four-torus.
The dual space is nothing but orbifold limit of the {\sl dual IIA} theory
that has been identified previously~\cite{schwarzsen}.

Incidentally, the
very same gauge theory is also obtained from the world-volume theory of
Type IIB NS five-brane~\cite{seiberg, wittennew, diaconescu}. There, the gauge coupling constant is independent
of string coupling parameter $\lambda_B$. In the limit $\lambda_B \rightarrow
0$, bulk dynamics decouples from the five-brane world-volume dynamics.

The theory consists of twisted sector as well.
On the parameter space, the orbifold fixed points are located at eight
fixed boundary circles. Located on each circle are (1+1)-dimensional
chiral fermions. In the bulk of the parameter space, the gauge theory
has a gauge group U(N)$\times$U(N) and the matter fields are all in
adjoint representation. As such, there is {\sl no} six-dimensional
gauge anomaly. What about at the boundary circles? At the boundary, 
there are zero-modes supported at the boundaries. These are cancelled
precisely by the twisted sector fermions. Again, we find that the
boundary (1+1)-dimensional gauge anomaly is cancelled. Incidentally,
through the Wess-Zumino consistency condition, the supersymmetry anomaly
is guaranteed to be absent if the gauge anomaly is cancelled.
It is precisely these twisted sector chiral fermions that describe the 
Wilson line moduli deformation of CHL string compactification.
The transverse positions of ${\bf S}_1$ circles on which the twisted sector 
chiral fermions wave function is localized can be deformed
continuously into the interior of the $\widetilde {\bf T}^4/\integer_2$.

\section{Discussion}
In this paper, we have studied M(atrix) theory description of 
M-theory compactified on unoriented two-dimensional manifold: Klein bottle
and M\"obius strip. The corresponding M(atrix) theories
are gauge theory of area-preserving diffeomorphism of Klein bottle and
M\"obius strip respectively. We have found that the M(atrix) theories are 
(2+1)-dimensional ${\cal N} = 8$ supersymmetric U(N) gauge theory. 
The parameter space on which each M(atrix) gauge theory lives is 
{\sl dual Klein bottle} or {\sl dual M\"obius strip} respectively. 

To illustrate how the result can be extended to higher-dimensional
non-orientable manifold, we have investigated strongly coupled limit
of toroidally compactified six-dimensional CHL heterotic string. 
We have shown that low-energy limit of corresponding M(atrix) theory
is six-dimensional gauge theory with gauge group U(N)$\times$U(N)
and with (1,0) supersymmetry. We have shown that the parameter space on
which the gauge theory lives is {\sl dual} $\widetilde{\bf S}_1 \otimes
\widetilde{K3} / \integer_2$, where the $\integer_2$ action acts as 
parity on $\widetilde{\bf S}_1$ and as a half-period shift on one of the
coordinates of $\widetilde{K3}$.
Seiberg has proposed that M(atrix) theory description of Type IIA on K3
is given by Neveu-Schwarz five-brane compactified on $\widetilde{K3}
\times {\bf S}_1$. We expect that the description extends to the present
situation, where K3 is modded out further by a freely acting involution
$\integer_2$. The M(atrix) theory description should then be 
in terms of Neveu-Schwarz five-brane on dual manifold $\widetilde{K3} \times
{\bf S}_1/ \integer_2$. In fact, the gauge theory we have deduced
corresponds to low-energy description of Neveu-Schwarz five-brane of Type 
IIB theory compactified on this dual space. Details of the investigation will 
be reported elsewhere.

\vskip0.5cm 
We are grateful to E. Witten for helpful discussions.


\end{document}